\documentclass{ifacconf}
\usepackage{graphicx}      
\usepackage{epstopdf}

\begingroup
\catcode`\#=11

\endgroup

\usepackage{multirow}
\usepackage{natbib}        
\usepackage{enumitem}
\usepackage{tabularx}
\usepackage{placeins}
\usepackage{algorithm,algpseudocode}

\algnewcommand{\Initialize}[1]{%
  \State \textbf{Initialize:}
  \Statex \hspace*{\algorithmicindent}\parbox[t]{.8\linewidth}{\raggedright #1}
}
\let\ifacconfcaptionwidth\captionwidth
\usepackage[caption=false]{subfig}
\let\captionwidth\ifacconfcaptionwidth

\usepackage{amsfonts,color}
\usepackage{amsmath}
\usepackage{accents}
\usepackage{mathrsfs}
\usepackage{amssymb, bm}

\def\Gin@extensions{%
          .png,.pdf,.jpg,.mps,.jpeg,.jbig2,.jb2,%
          .PNG,.PDF,.JPG,.JPEG,.JBIG2,.JB2,%
          .eps%
      }%
      
\DeclareMathOperator*{\argmin}{argmin} 

 \makeatletter
\newcommand{\closenomencl}{%
  \closeout\@nomenclaturefile%
}

\newtheorem{hyp}{\sc Assumption}
\makeatother
\begin{document}
\begin{frontmatter}

\title{Investigation of fast-NMPC and deep learning approach in fixed-point-based hierarchical control.} 


\author[First,Third]{Xuan-Huy Pham} 
\author[First]{Mazen Alamir} 
\author[Third]{François Bonne}

\address[First]{University of Grenoble Alpes, Gipsa-lab,  
   (xuan-huy.pham@grenoble-inp.fr).}
\address[Third]{Univ. Grenoble Alpes,IRIG-DSBT,
 F-38000, Grenoble,France}

\begin{abstract}                
This paper explores some variations of a hierarchical control framework that has been recently proposed. The framework is dedicated to control a network of interconnected subsystems such as the ones describing cryogenic processes or power plants. Recent investigations showed that handling constraints and nonlinearities might challenge the real-time feasibility of the approach. This paper investigates and combine two successful directions, namely, the use of truncated fast gradient and deep neural networks based controller modeling in order to reduce the computation time of the most critical subsystem. It is also shown that by doing so, the control updating period can be drastically reduced and the closed-loop performances highly improved. The paper can therefore be seen as a concrete implementation and validation of some key ideas in real-time distributed NMPC design. All the concepts are validated using the realistic and challenging example of real-life cryogenic refrigerator.
\end{abstract}


\end{frontmatter}

\section{Introduction}
In nuclear fusion reactors or particle accelerators, cryogenic refrigerators play a critical role as they cool down the thermal loads on the superconducting magnets to maintain the functionality of the overall process \citep{HENRY20071454,claudet2000economics}. These facilities are composed of several highly coupled subsystems forming an interactive network that requires efficient control design.

In this context, many studies have been conducted in the area of system modeling and model-based control methodology: In the work of \citep{Bonne2014ExperimentalIO} and references therein, several model-based multi-variable and constrained control strategies were investigated. Nevertheless, the aforementioned works are based on centralized frameworks which have some obvious drawbacks. Indeed, for such large systems where subsystems may be geographically located in different buildings, operators prefer modular design in order to facilitate testing, upgrading and maintenance operations. On the other hand, completely decentralized PID-based schemes fail to achieve an optimal design while satisfying the operating and safety constraints. 

Recently, a fixed-point-based hierarchical control framework has been suggested and validated for an interacting network of subsystems \citep{alamir2017fixed,Pham2021}. The proposed hierarchical control architecture is structured in two distinct layers. In the upper layer, a coordinator exchanges information with the subsystems located at the lower layer. By using the information exchanged with the subsystems, the coordinator minimizes a global cost by computing an optimal vector of set-points to be sent to the subsystems. On the other hand, each subsystem in the lower layer implements a local controller in order to regulate a specific output vector. While keeping the whole structure proposed in \citep{alamir2017fixed} unchanged, \citep{Pham2021} have demonstrated the validity of the methodology under constraints on actuators and nonlinearities in the underlying models. However, it pointed out that the relating computational burden might become a challenging issue since the  optimization control problems at the local level have to be solved repeatedly during a  fixed-point iteration loop. In order to address this problem, the authors in \citep{Pham2021} proposed to reduce the complexity of the optimizing set-point vector by performing the distribution of the optimization process over cyclically changed decision variables, aiming at limiting the number of iterations per updating period. However, this technique induces a slight drop in the resulting closed-loop performance. 

This paper attempts to address the same issue following the tracks described hereafter: \\ \ \\
(a) First it is shown that using a truncated fast gradient algorithm enables to reduce the computation time compared to generic available framework (such as CasADi/IPOPT \citep{Andersson2019CasADiAS}) even if the maximum number of iterations of such solver is limited (see Table \ref{table_compare_solver}). \\ \ \\
(b) Then, in order to further reduce  the computation time of the most cpu-critical local controller, a feed-forward deep neural network is used to approximate the control law provided by the fast gradient algorithm.\\ \ \\
(c) Finally, capitalizing on the resulting reduction of computation time, it is shown that by adopting a smaller control updating period, one can significantly improve the closed-loop performance. 

This paper is organized as follows: Section \ref{recall_hier} recalls the hierarchical control framework. Section \ref{FG_section} recalls the fast-gradient algorithm. Section \ref{DNN_section} describes the data generation used to train the deep neural network. Finally, section \ref{result_section} describes the illustrative example and shows numerical results to assess the effectiveness of proposed approaches. 

{\bf Notation}. Let $\oplus$ denotes the concatenation operator, namely for a sequence of vectors  $q_{i_1}, q_{i_2}, \dots$ define:
\begin{equation}
\underset{i \in \mathcal{I}}{\oplus}q_i : = [q_{i_1}^T,q_{i_2}^T, \dots]^T,\, \, \text{with} \quad i_1 < i_2 < \dots \in \mathcal{I}
\end{equation}
Moreover, the bold-faced notation $\bm p$ denotes the profile of a vector variable $p$ over a prediction horizon of length $N$, namely:
\begin{equation}
\bm p = [p^T(k), \dots, p^T(k+N-1)]^T \in \mathbb{R}^{N\cdot n_p}
\end{equation}

\section{Recall on fixed-point based hierarchical control framework:} \label{recall_hier}
\begin{figure}[h]
 \centering
 	\includegraphics[width = 0.7\linewidth]{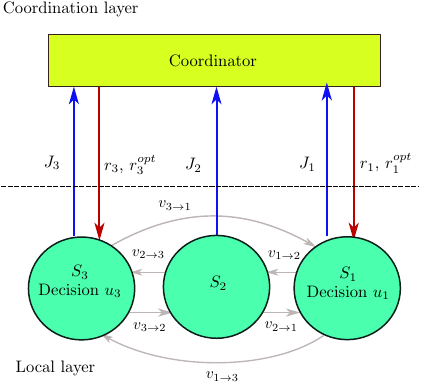}
 	\caption{Example of the hierarchical control architecture and the interconnection network between the subsystems. The presented sets correspond to this example are $\mathcal{N}:= \{1,2,3\}$; $\mathcal{N}^{ctr} := \{1,3\}$, $\mathcal{N}_1 = \{2,3\}$, $\mathcal{N}_2 = \{1,3\}$, $\mathcal{N}_3 = \{1,2\}$.} \label{Fig_inter}
 \end{figure}

Fig. \ref{Fig_inter} described the case of interest where a set of interacting subsystems indexed by $\mathcal{N}:= \{1,\dots,n_s\}$ is represented. This set is subdivided into two different subsets: 
\begin{itemize}
\item A subset of controlled subsystems indexed by $\mathcal N^{ctr}\subset \mathcal N$ having each its control input vector and regulated output vector, denoted for any $s\in \mathcal N^{ctr}$ by $u_s$ and $y_s$ respectively. 
\item A potential complementary subset of subsystems that includes no control input denoted by $N^{unc}:= \mathcal N-\mathcal N^{ctr}$.
\end{itemize}  
The dynamic of each subsystem $S_s$ is impacted through the so-called coupling signal $v_{s' \rightarrow s}$ coming from all exogenous subsystems  $\{S_{s'}\}_{s'\in \mathcal{N}_s}$ with indices $s'$ belonging to the set of indices $\mathcal N_s$ (set of indices of subsystems impacting $S_s$).

Let $\bm v^{in}_s$ and $\bm v^{out}_s$ be the incoming/outgoing coupling profiles into and from the subsystem $S_s$ respectively. More precisely:
\begin{equation}
\bm v_s^{in} :=   \underset{s' \in \mathcal{N}_s}{\oplus} \bm v_{s' \rightarrow s} \quad;\quad 
\bm v_s^{out} := \underset{s' \vert s \in \mathcal{N}_{s'}}{\oplus} \bm v_{s \rightarrow s'}
\end{equation}
In \citep{pham2021generic} the generic formulation of the proposed framework has been well defined.  However, it is essential to recall the overall hierarchical control framework. Let us begin with the following assumption:
\vskip 2mm
\hrule 
\vskip 2mm
\begin{hyp}
Each subsystem $S_s$, when given 
\begin{itemize}
\item a presumed incoming profile $\bm v_s^{in}$ and
\item a given individual set-point $r_s$ (required if $s\in \mathcal N^{ctr}$),
\end{itemize}   
can compute what would be:
\begin{itemize}
\item Its control profile $\bm u_s$ (if it has) by solving an nonlinear optimization problem,
\item Its resulting outgoing profile $\bm v_s^{out}$ and ,
\item Its contribution $J_s$ to the central cost.
\end{itemize}
The central cost is assumed to be of the form:
\begin{equation}
J_c(r, \bm v^{in}) := \sum_{s\in \mathcal N^{ctr}}J_s(r_s, \bm v_s^{in})+\sum_{s\in \mathcal N^{unc}}J_s(\bm v_s^{in})
\end{equation}  
where $r := \underset{s \in \mathcal{N}^{ctr}}{\oplus}r_s$ and $\bm v^{in} := \underset{s\in \mathcal N}{\oplus}\bm v_s^{in}$. Note that typical regulation-based cost $J_s$ is defined for $s\in \mathcal N^{ctr}$ while $J_s$ might represent a constraints violation indicator when $s\notin \mathcal N^{ctr}$.
\end{hyp}
\vskip 2mm
\hrule 
\vskip 2mm
More precisely, each time the coordinator sends $(r, \bm v^{in})$ to  the subsystems, this allows subsystems to compute (in parallel) their corresponding control profiles $\bm u_s$ (if $s\in\mathcal{N}^{ctr}$) and the outgoing coupling signal profiles $\bm v_s^{out}$ which is represented by the following form:
\begin{equation}
\bm v_s^{out} = g_{out}(r, \bm v^{in}) \label{vout_eq}
\end{equation}
Recall that both $\bm v^{in}$ and $\bm v^{out}$ are composed of all the elementary profiles $\bm v_{s \rightarrow s'}$. Hence, there is a matrix $G_{in}$ such that:
\begin{equation}
\bm v^{in} = G_{in} \cdot \bm v^{out} \label{vin_eq}
\end{equation}
By injecting  \eqref{vout_eq} into \eqref{vin_eq}, we obtain:
\begin{equation}
\bm v^{in} = G_{in} \cdot \bm g_{out}(r,\bm v^{in})
\end{equation}
Consequently, the problem that needs to be solved exclusively by the coordinator can be stated as follows:
\begin{align}
r^{opt} &= \argmin_r J_c(r,\bm v^{in}) \label{central_prob}\\
\text{subject to:} \, \bm v^{in} &= G_{in} \cdot \bm g_{out}(r,\bm v^{in})\label{v_cstr}
\end{align}
Since the coordinator does not have any mathematical knowledge of the subsystems the fixed-point map represented by \eqref{v_cstr} cannot be analytically known to the coordinator. That is the reason why the enforcement of \eqref{v_cstr} for a given set-point $r$ is done through a round of iterations between the coordinator and the subsystems as initially suggested in \citep{alamir2017fixed} where a fixed-point-iteration-based algorithm is proposed to evaluate a central cost associated to a given set-point $r$. Briefly, the algorithm could be summarized as below:
\begin{enumerate}
\item The coordinator starts by sending an initial guess $\bm v^{in,(\sigma = 0)}_s$ regarding  the incoming profiles,
\item The subsystems compute their control profiles (if any) and the corresponding in outgoing coupling profiles  $\hat{\bm v}^{out,(\sigma)}_s$ as well as their local cost $J_s$,
\item The subsystems send the outgoing coupling profiles  $\hat{\bm v}^{out,(\sigma)}_s$ to the coordinator from which the coordinator can constitute the corresponding incoming coupling profiles $\hat{\bm v}^{in,(\sigma)}_s$ based on \eqref{vin_eq}.
\item To ensure the convergence of the iteration, a stabilizing filter or a residual-based iterative method is used to update the profile denoted by $\bm  v^{in,(\sigma+1)}_s$,
\item The iterations continue until  the termination criteria $\epsilon := \text{max}(|\bm v^{in,(\sigma+1)}- \bm v^{in,(\sigma)} |) \leq $ is satisfied.  
\end{enumerate}
Having the cost associated to a given set-point, any derivative-free optimization algorithm can be used to solve \eqref{central_prob} in the decision variable $r$ (e.g. Genetic algorithm \citep{thede2004introduction}, BOBYQA \citep{Powell2009TheBA}, etc.). \section{Fast gradient method for solving NMPC problem} \label{FG_section}
Since the solution of the local NMPC problems is processed in parallel, the real-time feasibility of the  framework depends considerably on the critical subsystem requiring the highest computation time. Although many toolkits for solving optimization-based control problems, such as ACADO \citep{Houska2011ACADOTO} or CasADi \citep{Andersson2019CasADiAS}, it has been shown in \citep{alamir2014fast} that when a limited (computation time)/(hardware performance) are present, a truncated fast gradient might be beneficial to closed-loop performances. That is why this algorithm is briefly recalled here as it is in the heart of the forthcoming development. 

Recall that each subsystem $S_s$, $s\in \mathcal N^{ctr}$ solves an optimization problem upon receiving a pair of $(r_s,\bm v^{in}_s)$ from the coordinator:
\begin{align}
\mathcal{P}_s: \min_{\bm u_s \in \mathcal{U}_s} J^{loc}_s(\bm u_s)  \label{local_prob}
\end{align}
where $\mathcal{U}_s$ are the admissible set of control profiles $\bm u_s$. Note that the cost function $J^{loc}_s$ implicitly depends on the current state, the set-point $r_s$ and the incoming coupling profile $\bm v_s^{in}$. These variables are considered frozen during the solution of \eqref{local_prob} and are dropped for reasons of compactness. 

The implementation of the fast gradient method requires the gradient of the cost function at $J^{loc}_s$ with respect to $\bm u_s$, which can be easily obtained by modeling the cost with CasADi and then computing its gradient $\nabla J^{loc}_s$. The algorithm that is used to solve \eqref{local_prob} is given by the following updating rule:
\begin{align}
\bm z_s^{i+1} &= \bm u_s^i - \gamma \cdot \nabla J^{loc}_s(\bm u_s^i) \\
\bm u_s^{i+1} &= \textbf{Pr}(\bm z_s^{i+1} + c \cdot (\bm z_s^{i+1}- \bm z_s^i),\mathcal{U}_s)
\end{align}
where $c \in (0,1)$ is the design variable and $\textbf{Pr}(p,\mathcal{U}_s)$ is the projection of vector $p$ on the admissible set $\mathcal{U}_s$. The variable $\gamma$ is the step size that is calculated by using Barzilai-Borwein formula proposed in \citep{barzilai1988two}:
\begin{equation}
\gamma^{i+1} = \frac{\|(\bm u_s^{i+1}- \bm u_s^{i}) \cdot (\nabla J^{loc}_s(\bm u_s^{i+1}) -\nabla J^{loc}_s(\bm u_s^i))\|}{\|\nabla J^{loc}_s(\bm u_s^{i+1}) -\nabla J^{loc}_s(\bm u_s^i) \|^2} \label{gamma_eq}
\end{equation}
In \citep{10.2307/2156099}, the author shows that the convergence of the algorithm could be improved when a restart mechanism is included. More precisely, the variable $\bm u_s$ is restarted every $n_{rst}$ iteration, but it is noted that the frequency of restarts should depend on the cost function.

Finally, this method is summarized by Algorithm \ref{alg:1}.
\begin{algorithm}
\caption{Fast conjugate gradient method}\label{alg:1}
\begin{algorithmic}[1]
\Initialize{$i \leftarrow 0$; $c \in (0,1)$; $\gamma^i \in (0,1)$; $n_{rstr} \in \mathbb{N}$\\ $\bm u_s^i \leftarrow \textbf{0}$; $\bm z_s^i \leftarrow \textbf{0}$}\\
\For{$i \leftarrow 1,\dots,N_{max}$} 
\State $\bm z_s^{i+1} = \bm u_s^i - \gamma^i \cdot \nabla J^{loc}_s(\bm u_s^i)$;
\If{$\text{mod}(i,n_{rstr}) == 0$} \Comment{check for restart}
\State $\bm u_s^{i+1}  =  \textbf{Pr}(\bm z_s^{i+1},\mathcal{U}_s)$;
\Else{}
\State $\bm u_s^{i+1} = \textbf{Pr}(\bm z_s^{i+1} + c \cdot (\bm z_s^{i+1}- \bm z_s^i),\mathcal{U}_s)$;
\EndIf
\State Compute $\gamma^{i+1}$ by \eqref{gamma_eq};
\EndFor
\end{algorithmic}
\end{algorithm}

\section{Approximate NMPC by neural network} \label{DNN_section}
Recently, deep neural networks (DNNs) have become a popular choice for the functional form $K_{NN}(z;\theta)$ (with $z$ being the argument of the control law) because of their universal approximation property. Furthermore, DNNs could be easily implemented in any programmable logic controllers (PLCs), whose computational capabilities are not suitable for high-level solver. With $L$ hidden layers and H nodes per layer, a DNN is given by:
\begin{equation}
K_{NN}(z;\theta) = \alpha_{L+1} \circ \beta_{L} \circ \alpha_L \circ \dots \circ \beta_1 \circ \alpha_1(z)
\end{equation}
Each hidden layer involves affine transformation of the output of its previous layer:
\begin{equation}
\alpha_l(\epsilon_{l-1})= W_l \cdot \epsilon_{l-1} + b_l
\end{equation} 
in which $\epsilon_{l-1} \in \mathbb{R}^H$ for $l \in \{2,\dots, L+1\}$ and $\epsilon_0 = z$. The function $\beta_l$, for $l \in \{1,\dots,L\}$ are nonlinear activation functions (e.g, rectified linear units (Relu), sigmoid,...). The parameter vector $\theta = \{W_1,b_1,\dots, W_{L+1}, b_{L+1}\}$ gathers all weights $W_l$ and biases  $b_l$ in the network  with appropriate dimension
Once the network architecture is trained according to , the approximate DNN-based NMPC law $K_{NN}(z,\theta)$ can be used online to cheaply evaluate the optimal control input.
\subsection{Data generation}
There are two common data-generation strategies, namely open-loop and closed-loop. In open-loop data generation, the set $\mathcal{Z} \subset \mathcal{X}_s\times \mathcal{V}^{in} \times \mathcal{R}_s \times \mathcal{W}_s$ of possible states, incoming coupling profiles, disturbances and set-points could be created and the corresponding control profile $\bm u_s$ computed that will be added together to establish a set of data $\mathcal{D} = \{(x^{(i)}_s,  \bm v^{in,(i)}_s, r^{(i)}_s, \bm w^{(i)}_s, \bm u^{(i)}_s)\}_{i=1}^{N_s}$. Although very simple, this strategy can result non physically realistic instances being included in the training data. Closed-loop strategy, on the contrary, gathers data while running a closed-loop simulation under randomly drawn physically meaningful initial states. Indeed, the majority of large-scale cryogenic systems operate under a relatively small number of regimes or operating scenarios. Each operational scenario is characterized by a few controlled outputs and a few large magnitude disturbances that may change frequently, while the set-points are kept unchanged for a long period of time. Hence, we propose the following data generation procedure that performs off-line simulation using model to collect the operationally relevant training set $\mathcal{D}$:
\begin{enumerate}
\item Determine the operational range of the set-points denoted by $[\underline{r}_s, \overline{r}_s]$ and the realistic range of the disturbances denoted by $[\underline{w}_s, \overline{w}_s]$:
\item Create pseudo random binary signals (PRBS) of $r_s$ and $w_s$ in their operational ranges.
\item Run the closed-loop simulations that implement the above discussed hierarchical design at some chosen initial states with the created PRBS signals. Note that Data is collected during the fixed-point iterations in order to capture the relationship between the control profile $\bm u_s$ and the triplet $(r_s, x_s,\bm v^{in}_s$).
\end{enumerate}

The network is trained to minimize the mean squared error criteria below:
\begin{equation}
J_{NN}(\theta) = \frac{1}{2}\sum_{j=1}^{N_{tr}} [K_{NN}(z_s^{(j)},\theta) - \bm u_s^{(j)}]^2
\end{equation}
where $N_{tr}$ is the number of training observations. The resilient back-propagation (RPROP) algorithm is used to train the neural network. The activation function at each node is the sigmoid function. Many configurations of NN will be examined in the simulation section.\\
\section{System description / numerical results}\label{result_section}
\subsection{System description}
\begin{figure}[h]
 \centering
 	\includegraphics[]{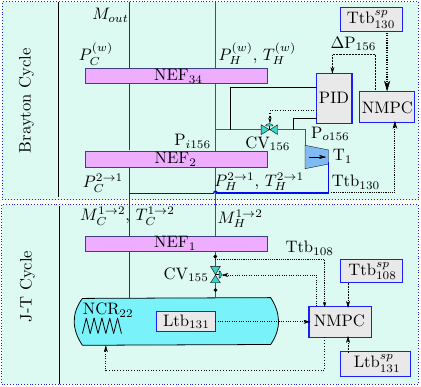}
 	\caption{Block diagram of the cold box plant.} \label{a}
 \end{figure}

\begin{figure}[h]
 \centering
 	\includegraphics[]{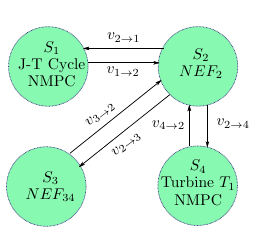}
 	\caption{Block diagram of the cold box plant.} \label{decomp_4}
 \end{figure}

The system under investigation is the cold box of a cryogenic refrigerator (Fig. \ref{a}) composed of a Joule-Thomson cycle and a Brayton cycle. The Brayton cycle is composed of two heat exchangers, which are NEF$_2$, NEF$_{34}$ and a cryogenic turbine T$_1$. The thermal energy from the helium flow is extracted by using the turbine T$_1$ and by exchanging the heat power between high pressure pipe line and low pressure pipe line through a series of heat exchangers (NEF$_x$). When passing the valve CV$_{155}$ (Joule-Thomson cycle), the isenthalpic process occurs, resulting in the liquefaction of part of the gaseous helium, which rests in the helium bath. The remaining gaseous part return to the cycle through the low pressure line.\\ \ \\
\textbf{The Manipulated Inputs:} There are three control inputs which are CV$_{155}$, NCR$_{22}$ belonging to Joule-Thomson cycle and $\Delta \text{P}_{156}$ which is a part of the Brayton cycle. These actuators are defined below:
\begin{enumerate}
\item CV$_{155} \in [0\%,100\%]$: This valve is situated at the inlet of the helium bath. 
\item NCR$^{(a)}_{22}$: This heating actuator is located inside the helium bath. The value of $\text{NCR}_{22}^{(a)}$ is in the range of $[0,55]$ W. Note that the variable NCR$_{22}$ in Fig. \ref{a} is decomposed into two terms:
\begin{equation}
\text{NCR}_{22} := \text{NCR}_{22}^{(a)} + \text{NCR}_{22}^{(w)}
\end{equation}
where $\text{NCR}_{22}^{(w)}$ represents the disturbance coming from the heat source.
\item $\Delta \text{P}_{156} \in  [0,12]$ bar: The pressure drop between the inlet pressure and outlet pressure of the valve CV$_{156}$. Indeed, in order to hide the nonlinearity, the valve CV$_{156}$ is controlled through the pressure drop between its inlet and outlets. Precisely, the required pressure drop is calculated by an NMPC and sent to the PID controller, which acts on the opening position of the CV$_{156}$ valve (Fig. \ref{a}).
\end{enumerate}
\textbf{The Regulated Outputs:} There are three regulated outputs and one constrained output (Figure \ref{a}):
\begin{enumerate}
\item Ltb$_{131}$: The helium liquid level (\%) that must be controlled to ensure that some thermal loads are always extracted (e.g. used to cool super-critical helium at liquid helium temperature to be ready for the final customer). The set-point is chosen by the operator. In the usual operation, it is set at Ltb$_{131}^{sp}= 60.5\%$.
\item Ttb$_{108}$: The temperature at the inlet of the J-T valve must be tightly controlled in order to ensure the efficiency of the liquefaction of the helium.
\item Ttb$_{130}$: Since the cryogenic turbine is a critical component, the temperature at its outlet must be tightly regulated to avoid the risk of liquid droplet forming at the outlet, potentially destructive for the turbine.
\item $M_{out}$: The exiting flow rate of exchanger NEF$_{34}$ is constrained to be lower than an allowed maximum flow rate $\overline{M}_{out} = 0.07$ kg/s.
\end{enumerate} 
This system could be viewed as a network of four coupled subsystems (Fig. \ref{decomp_4}). 
Table \ref{table_network} summarizes the inputs and outputs of each subsystems in this decomposition. The notation $T_C$, $M_C$, and $P_C$ ($T_H$, $M_H$, and $P_H$) are respectively the temperature, flow rate, and pressure of the cold (hot) branch of the refrigerator.
\begin{table}[h]
\centering
\caption{The inputs, outputs and the coupling variables of the 4-subsystems topology.} \label{table_network}
\renewcommand*{\arraystretch}{1.3}
\begin{tabular}{c|c|c|l}
\hline
       &   $u_s$      & $y_s$    & \multicolumn{1}{c}{$v_{s \rightarrow s'}$}     \\ \hline                                                                                                                                                                                          
$S_{1}$  &\begin{tabular}[c]{@{}c@{}}NCR$_{22}^{(a)}$\\ CV$_{155}$\end{tabular}   & \begin{tabular}[c]{@{}c@{}}Ltb$_{131}$\\ Ttb$_{108}$\end{tabular} & $v_{1 \rightarrow 2} = [M_{H}^{1 \rightarrow 2},M_{C}^{1 \rightarrow 2},T_{C} ^{1 \rightarrow 2}]^T$              \\ \hline
$S_{2}$       &  \_          & \_     & \begin{tabular}[c]{@{}l@{}l@{}} $v_{2 \rightarrow 1} = [T_{H}^{2 \rightarrow 1}, P_{H}^{2 \rightarrow 1}, P_{C}^{2 \rightarrow 1}]^T$ \\  $v_{2 \rightarrow 3} = [M_H^{2 \rightarrow 3}, M_C^{2 \rightarrow 3}, T_C^{2 \rightarrow 3}]^T$ \\  $v_{2 \rightarrow 4} = [P_C^{2' \rightarrow 4}]$ \end{tabular} \\ \hline
$S_{3}$       &\_             &  $M_{out}$                                                              & \begin{tabular}[c]{@{}l@{}}$v_{3 \rightarrow 2} = [T_{H}^{3 \rightarrow 2}, P_H^{3 \rightarrow 2}, P_C^{3 \rightarrow 2}]^T$\\ $v_{3 \rightarrow 4} = [T_{H}^{3 \rightarrow 4},P_H^{3 \rightarrow 4}]^T$\end{tabular}            \\ \hline
$S_{4}$    &     $\Delta$P$_{156}$         & Ttb$_{130}$                                                    & \begin{tabular}[c]{@{}l@{}}$v_{4 \rightarrow 2} = [M_C^{4 \rightarrow 2}, T_C^{4 \rightarrow 2}]^T$\\ $v_{4 \rightarrow 3} = [M_H^{4 \rightarrow 3}]$\end{tabular}                                                             \\ \hline
\end{tabular}
\end{table}

The following local costs of each subsystem are used:
For $S_{1}$ and $S_{4}$ that need to track the desired set-point $r^d_s$:
\begin{equation}
J_s(r|r^d_s)= \sum_{i=0}^{N-1} \| y_s(k+i)-r^d_s \|^2_{Q_c^{(s)}} + \| u_s(k+i) \|^2_{R_c^{(s)}} \label{cost14}
\end{equation}
For $S_{3}$ that has output to be constrained
\begin{equation}
J_{3}(r|\overline{y}_3)= \sum_{i=0}^{N-1} \|\max(y_{3}(k+i)-\overline{y}_{3},0)\|^2_{Q_c^{(3)}}\label{cost3}
\end{equation}
Finally, $S_{2}$ does not have any contribution to the central cost, its cost is simply defined by $J_{2}(r) = 0$. Note however that this subsystem impacts the fixed-point definition as its outlet $\bm v_s^{out}$ depend on the incoming parameters. The weighing matrices appears in \eqref{cost14}-\eqref{cost3} is listed below:
\begin{align}
Q_c^{(1)} &= \text{diag}([10^3,10^3]) &&\quad Q_c^{(4)} = 10^3  & Q_c^{(3)} &= 10^{10}\\
R_c^{(1)} &= \text{diag}([0,0]) &&\quad R_c^{(4)} = 0
\end{align}
where $\text{diag}()$ denotes a diagonal matrix. 

In order to compare the performance of two strategies, the closed-loop performance $J_{c}^{cl}$ is used that is defined by:
\begin{align}
J_{c}^{cl} = \frac{1}{N_{sim}} \sum_{i=1}^{N_{sim}} \sum_{s \in \mathcal{N}} J_s^{cl}(i)
\end{align}
where $N_{sim}$ is the simulation time duration , $J_s^{cl}(i)$ is computed according to the criteria of each subsystem as defined in \eqref{cost14}-\eqref{cost3}
 
Since the local controller of the Joule-Thomson cycle ($S_1$)  has the most critical computation time compared to one of the Brayton cycle, it will be approximated by the DNN. Concerning the data generation to train the DNN, the procedure described in section \ref{DNN_section} is performed while optimal control profiles are obtained by using the fast gradient method because the computation time of CasADi (which is 0.5 secs for resolving problem \eqref{local_prob} is highly impractical to be used in this framework which will be shown in the beginning of subsection \ref{result_section}. 

\subsection{DNN model assessment}
The learning performance is evaluated for three different configurations of DNNs. These configurations are set up so that each DNN has a different number of hidden layers, ranging from 1 to 3 layers, with each layer having the same number of nodes, i.e. 25 nodes, denoted by NN-1-25, NN-2-25 and NN-3-25, respectively. Concretely, each structure is trained for $10000$ epochs, with $450\times 10^3$ samples and validated with $450\times 10^3$ in the same sample pool. Table \ref{MSE_DNN} presents the learning performance for three DNN structures. The structure NN-2-25 which has the lowest MSE is chosen to conduct the next simulation. 

\begin{table}[h]
\centering
\caption{The learning performance of several configuration of DNNs.} \label{MSE_DNN}
\begin{tabular}{c|c|c|c}
\hline
Structure  & NN architecture      & MSE   & Training time \\ \hline
NN-1-25   &  [25 25 12]  & 0.3192 &   2h47 \\
NN-2-25  &  [25 25 25 12] & 0.2726 &   3h15  \\
NN-3-25 &  [25 25 23 25 12]& 0.2996 &   3h50 \\ \hline
\end{tabular}
\end{table}
\subsection{Numerical result} \label{result_section_really}
First, we compare the performance of the Ipopt (CasADi) solver and the truncated gradient solver used to solve the problem \eqref{local_prob} of $S_1$. This can be done by evaluating the open-loop performance indicated by $J_s^{loc}(\bm u_s^*)$, where $\bm u_s^*$ is the solution of \eqref{local_prob}. The evaluation process is described below:
\begin{enumerate}
\item Create realistic set of state $x_s$, set-point $r_s$ and  $\bm v_s^{in}$ denoted by $\mathcal{D}^{\text{solver}}:= \{(x_s^{(i)},r_s^{(i)},\bm v_s^{(i)})\}_{i=1}^{N_{dta}}$;
\item Solve the problem \eqref{local_prob} by using solver Ipopt and truncated gradient at triplets $(x_s^{(i)},r_s^{(i)},\bm v_s^{(i)})$ (for $i = 1,\dots,N_{dta}$);
\item The open-loop performances $J_s^{loc,(i)}(\bm u_s^{*,\text{Ipopt}})$ and  $J_s^{loc,(i)}(\bm u_s^{*,\text{grd}})$ of the solver Ipopt and truncated gradient are computed. Then, the average of performance ratio $\overline{J}$ between the two solvers is computed, namely:
\begin{equation}
\overline{J} =  \frac{1}{N_{dta}}\sum_{i=1}^{N_{dta}} \frac{J_s^{loc,(i)}(\bm u_s^{*,\text{grd}})}{J_s^{loc,(i)}(\bm u_s^{*,\text{Ipopt}})} \times 100\%
\end{equation}
\end{enumerate}

The maximum number of iterations $N_{\text{iter}}^{\text{max}}$/ the acceptable tolerance $\epsilon_{\text{tol}}$ of  solver Ipopt is set at $5$ and $10^{-1}$, respectively. For the truncated gradient solver, the maximum number of iteration is set to $N_{max}= 50$ and the update variable is restarted at every $n_{rstr} = 5$ iterations.
\begin{table}[]
\caption{Performance index of solver Ipopt and truncate gradient.} \label{table_compare_solver}
\centering
\begin{tabular}{c|c|c|c|c}
\hline
                   & $N_{max}$ & $\epsilon_{\text{tol}}$ & $\overline{J}$ & $t^{\text{max}}_{\text{cpt}}$ \\ \hline
Truncated gradient & 50        & -                       & $99.81\% $       & 0.018s                        \\ \hline
IPOPT              & 5         & $10^{-1}$               & $100\%  $        & 0.502s                        \\ \hline
\end{tabular}
\end{table}
Table \ref{table_compare_solver} shows the average performance ratio $\overline{J}$ and the maximal computation time $t^{\text{max}}_{\text{cpt}}$ associated to the two solvers. It can be noted that the truncated gradient solver gives a more far lower time while achieving in average the same open-loop performance compared to those given by Ipopt solver.

\begin{figure}[h]
 \centering
  \includegraphics[trim= {1.5cm 10cm 1.5cm 9.25cm},width =\linewidth, clip]{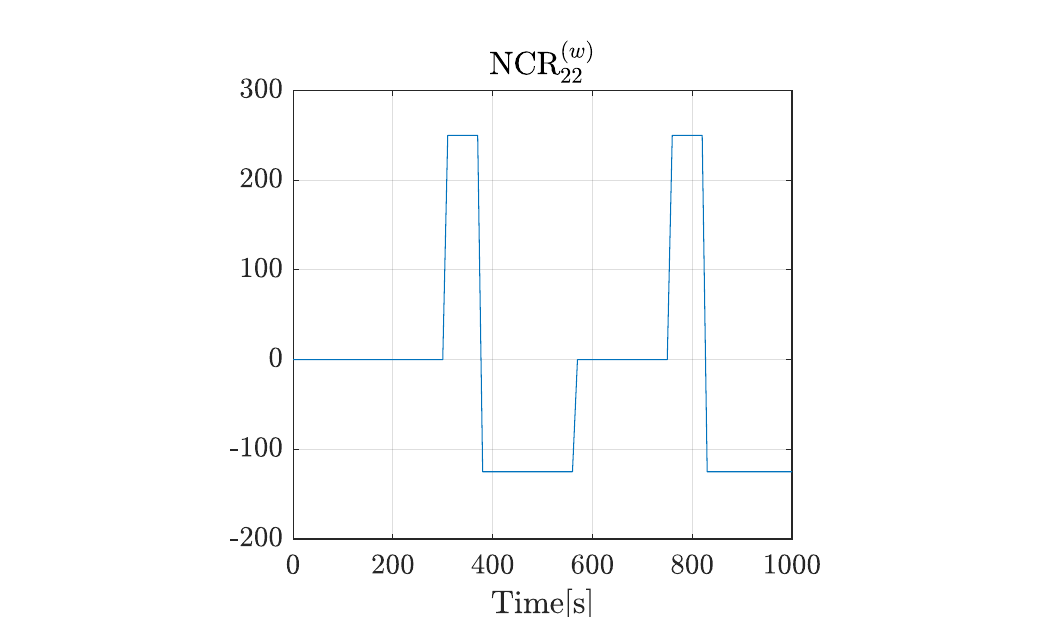} 
  \caption{Disturbance profile.} \label{dis_pro}  
\end{figure}

\begin{figure}[h]
 \centering
  \includegraphics[trim= {1.5cm 4.5cm 1.5cm 5cm},width =\linewidth, clip]{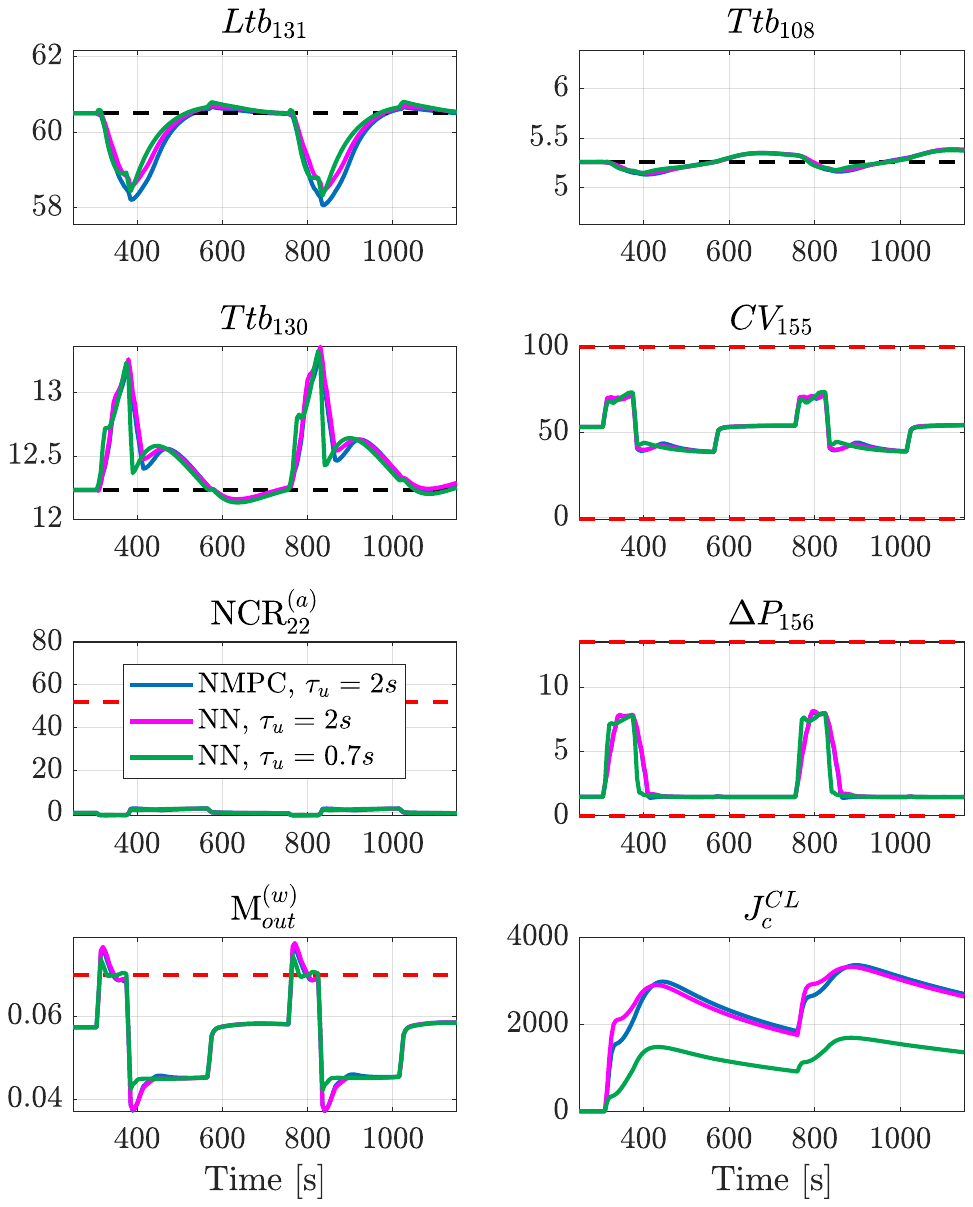} 
  \caption{Comparison of the system behavior with different configurations.} \label{compare_sp}  
\end{figure}

Second, we compare the performance of the hierarchical control framework with different set-ups. More precisely, we compare the closed-loop performance indices $J_c^{CL}$s under the disturbance profile (Fig. \ref{dis_pro}) when using the exact NMPC of the Joule-Thomson cycle, the approximation NN-2-25 with $\tau_u = 2s$ and with $\tau_u = 0.7s$ (Fig. \ref{compare_sp}). It can be noted that the DNN approach allows to update more frequently the control decision, which results a better performance in closed-loop. 

\begin{figure}[h]
 \centering
  \includegraphics[trim= {1.2cm 8.5cm 1.1cm 9cm},width =\linewidth]{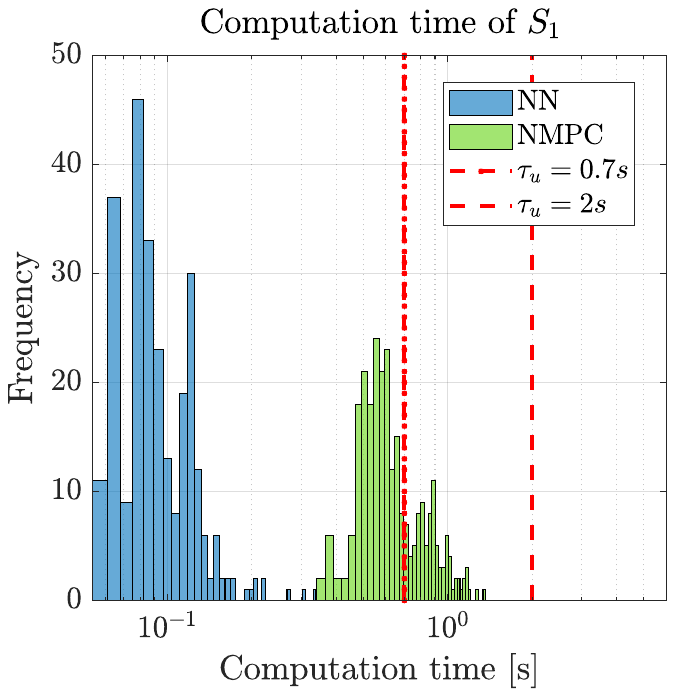} 
  \caption{Computation time of NN-2-25 and NMPC.} \label{compare_time}  
\end{figure}

Finally, Fig. \ref{compare_time} shows the computation time associated to the Joule-Thomson cycle within the hierarchical control framework.
\section{Conclusion}
In this paper, two methods have been proposed to reduce the computation time of solving the constrained nonlinear optimization problem at the local layer of the hierarchical control framework. The numerical results have demonstrated the effectiveness of the two methods. On going work aims to validate the control structure with a full cryogenic facility.

\bibliography{mybibfile_2}             
                                                   







\end{document}